\documentclass[nonacm,sigconf]{acmart}
\usepackage{ascmac}
\usepackage{booktabs}
\usepackage{subfigure}
\usepackage[norelsize,ruled,vlined,linesnumbered]{algorithm2e}
\usepackage{algorithmic}

\newtheoremstyle{mystyle} 
    {1.5mm}
    {1.5mm}
    {\it}
    {0mm}
    {\scshape}
    {.}
    { }
    {}
\theoremstyle{mystyle}

\newtheorem{definition}{Definition}
\newtheorem{example}{Example}
\newtheorem{lemma}{Lemma}
\newtheorem{theorem}{Theorem}
\newtheorem{corollary}{Corollary}

\newcommand{\wsq}{\hspace{\fill}$\square$}
\newcommand{\vs}{\vspace{1.5mm}}
\SetKwComment{Comment}{$\triangleright$\ }{}

\AtBeginDocument{%
  }

\setcopyright{rightsretained}
\copyrightyear{2023}
\acmYear{2023}
\acmDOI{10.1145/3603719.3603720}
\acmISBN{979-8-4007-0746-9/23/07}
\acmConference[]{}{}{}
\acmBooktitle{}

\makeatletter
\gdef\@copyrightpermission{
  \begin{minipage}{0.3\columnwidth}
   \href{https://creativecommons.org/licenses/by-nc/4.0/}{\includegraphics[width=0.90\textwidth]{figure/4ACM-CC-by-nc-88x31.eps}}
  \end{minipage}\hfill
  \begin{minipage}{0.7\columnwidth}
   \href{https://creativecommons.org/licenses/by-nc/4.0/}{This work is licensed under a Creative Commons Attribution-NonCommercial International 4.0 License.}
  \end{minipage}
  \vspace{5pt}
}
\makeatother

\setcopyright{none}
\renewcommand\footnotetextcopyrightpermission[1]{} 
\begin{document}

\title{Efficient Algorithms for Top-k Stabbing Queries on Weighted Interval Data (Full Version)}

\author{Daichi Amagata}
\affiliation{%
  \institution{Osaka University}
  \country{Japan}
}
\email{amagata.daichi@ist.osaka-u.ac.jp}

\author{Junya Yamada}
\affiliation{%
  \institution{Osaka University}
  \country{Japan}
}
\email{yamada.junya@ist.osaka-u.ac.jp}

\author{Yuchen Ji}
\affiliation{%
  \institution{Osaka University}
  \country{Japan}
}
\email{ji.yuchen@ist.osaka-u.ac.jp}

\author{Takahiro Hara}
\affiliation{%
  \institution{Osaka University}
  \country{Japan}
}
\email{hara@ist.osaka-u.ac.jp}

\begin{abstract}
Intervals have been generated in many applications (e.g., temporal databases), and they are often associated with weights, such as prices.
This paper addresses the problem of processing top-k weighted stabbing queries on interval data.
Given a set of weighted intervals, a query value, and a result size $k$, this problem finds the $k$ intervals that are stabbed by the query value and have the largest weights.
Although this problem finds practical applications (e.g., purchase, vehicle, and cryptocurrency analysis), it has not been well studied.
A state-of-the-art algorithm for this problem incurs $O(n\log k)$ time, where $n$ is the number of intervals, so it is not scalable to large $n$.
We solve this inefficiency issue and propose an algorithm that runs in $O(\sqrt{n}\log n + k)$ time.
Furthermore, we propose an $O(\log n + k)$ algorithm to further accelerate the search efficiency.
Experiments on two real large datasets demonstrate that our algorithms are faster than existing algorithms.
\end{abstract}

\maketitle

\section{Introduction}  \label{sec:introduction}
Many applications deal with interval data, where an interval is a pair of left and right endpoints.
For example, objects associated with time information (e.g., sales items and vehicles) are usually maintained in interval format (e.g., the left and right endpoints are activation and termination time, respectively \cite{amagata2017mining,amagata2016sliding,nishio2022lamps,behrend2019period,christodoulou2022hint,christodoulou2023hint}).
In cryptocurrency and stock applications, the prices of cryptocurrencies and stocks vary continuously, and they record minimum and maximum prices (i.e., an interval) every certain time \cite{qiao2016range,zhang2022approximate}.
It is also intuitively known that each interval usually has a weight \cite{agarwal2005optimal,kaplan2003dynamic}.
For instance, in the sales items and vehicles examples, the weights can be profits and the number of passengers, respectively.

\subsection{Motivation and Challenge}
To analyze the above weighted interval data, the following example queries can be considered:

\vs
\noindent
$\bullet\,$ Show $k$ vehicles (e.g., trains) with the largest number of passengers at noon yesterday.

\vs
\noindent
$\bullet\,$ Show $k$ intervals with the largest values of $(\max - \min)$ among a set of intervals including my buying price (e.g., in a cryptocurrency dataset).

\vs
\noindent
The first query helps consider a train operation plan and analyze train usage patterns for some events that occurred at a certain time.
The other query can find price increase patterns to obtain profits.
Motivated by these applications and usefulness, we address the problem of processing top-k weighted stabbing queries on interval data.
Note that, because a simple stabbing query does not consider weights and returns all stabbed intervals, applications cannot control the result size.
That is, they may be overwhelmed by large result sizes, so the controllable result size (i.e., the top-k factor) is useful for such applications.

Given a set $X$ of $n$ weighted intervals and a query $q = (s,k)$ where $s$ and $k$ are respectively a query value and a result size, this query retrieves $k$ intervals \textit{stabbed by $s$} with the largest\footnote{Some applications may prefer smaller weights, and our algorithms can deal with this case.} weight among $X$.
Note that an interval $x \in X$ is stabbed by $q$ iff $s \in [x.l, x.r]$, where $x.l$ and $x.r$ are the left and right endpoints, respectively.
Because many applications deal with large sets of intervals (i.e., $n$ is large), an efficient algorithm for this problem is required.
However, designing such an algorithm is non-trivial and challenging.

The most straightforward algorithm is as follows.
We sort the intervals $\in X$ in descending order of weight offline.
Given a top-k weighted stabbing query, we run a sequential scan of $X$ until we access $k$ stabbed intervals.
Due to the sort order, (i) this set of the $k$ intervals is guaranteed to be the exact top-k result, and (ii) this algorithm can stop the scan before accessing $n$ intervals.
However, in the worst case, this algorithm needs to access all intervals, so it incurs $O(n\log k)$ time.
(The factor of $O(\log k)$ is required to update the intermediate top-k result.)
Another approach is to employ a state-of-the-art algorithm \cite{xu2017efficiently}.
This algorithm uses an interval tree \cite{edelsbrunner1980dynamic} to find all stabbed intervals, and the top-k intervals are found from them.
Because the interval tree structure guarantees that a (non top-k weighted) stabbing query can run in $O(\log n + m)$ time, where $m$ is the number of stabbed intervals, this algorithm can run in $O(\log n + m\log k)$ for our problem.
At first glance, this algorithm seems sufficiently fast, but it is important to notice that $m$ can be as large as $n$ (e.g., all intervals are stabbed by a query).
Therefore, this algorithm results in the same worst time as the sequential scan.

\begin{table*}[!t]
    \centering
    \caption{Time complexity of each algorithm, where $n$ ($m$) is the number of (stabbed) intervals, and $m$ can be $O(n)$.}
    \label{tab:time}
    \begin{tabular}{cccc}      \toprule
        Algorithm                               & Pre-processing            & Query                     & Space             \\ \midrule
        Sequential scan                         & $O(n\log n)$              & $O(n\log k)$              & -                 \\ 
        Interval tree \cite{xu2017efficiently}  & $O(n\log n)$              & $O(\log n + m\log k)$     & $O(n)$            \\ 
        Segment tree                            & $O(n\log n)$              & $O(\log n + m\log k)$     & $O(n\log n)$      \\ 
        Ours-1                                  & $O(n\log n)$              & $O(\sqrt{n}\log n + k)$   & $O(n)$            \\ 
        Ours-2                                  & $O(n\log n\log\log n)$    & $O(\log n + k)$           & $O(n\log^{2}n)$   \\ \bottomrule
    \end{tabular}
\end{table*}

\subsection{Contribution}
The existing techniques suffer from $O(n\log k)$ time.
We hence have a question:
For our problem, does there exist an exact algorithm with less than $O(n)$ query time (and with $\tilde{O}(n)$ space, where $\tilde{O}(\cdot)$ hides any polylog factors)?
We provide a positive answer and make the following contributions\footnote{This is a full version of \cite{amagata2024efficient}.}:

\vs
\noindent
\underline{$\bullet\,$ \textit{An $O(\sqrt{n}\log n + k)$ time algorithm} (Section \ref{sec:intervalforest}).}
We first propose an algorithm that exploits weight-based sorting and the interval tree structure.
This technique provides a performance guarantee dominating that of the state-of-the-art algorithm \cite{xu2017efficiently}, because our algorithm runs faster than the state-of-the-art with the same space requirement.
As $\sqrt{n}\log n < n$, we have $O(\sqrt{n}\log n + k) < O(n)$.

\vs
\noindent
\underline{$\bullet\,$ \textit{An $O(\log n + k)$ time algorithm} (Section \ref{sec:stpsa}).}
The second algorithm improves the search efficiency by exploiting the segment tree structure \cite{de2000computational}.
A segment tree yields the same performance for simple stabbing queries, i.e., its time complexity is $O(\log n + m)$, so simply applying this structure still incurs $O(n\log k)$ time in the worst case.
Nevertheless, we show that a simple modification of this structure provides an $O(k\log n)$ time algorithm for our problem.
We furthermore extend the segment tree to reduce the time complexity from $O(k\log n)$ to $O(\log n + k)$.
Table \ref{tab:time} compares our new theoretical results with those of the existing techniques for top-k weighted stabbing queries.

\vs
\noindent
\underline{$\bullet\,$ \textit{Experiments on real datasets} (Section \ref{sec:experiment}).}
We conduct experiments on two real large datasets.
One has a small $m$, whereas the other has a large $m$.
In both cases, our algorithms outperform the existing algorithms.
Moreover, our $O(\log n + k)$ time algorithm requires \textit{only less than two microseconds} for $k \in [25, 100]$.

\vs
\noindent
In addition to the above contents, Section \ref{sec:preliminary} formally defines the problem addressed in this paper and introduces preliminary information.
Related works are reviewed in Section \ref{sec:related-work}, and finally, we conclude this paper in Section \ref{sec:conclusion}.

\section{Preliminary}   \label{sec:preliminary}

\subsection{Problem Definition} \label{sec:problem-definition}
We use $X$ to denote a set of $n$ intervals.
Each interval $x \in X$ is a pair of its left and right endpoints, i.e., $x = [x.l, x.r]$, where $x.l \leq x.r$.
In addition, each interval $x \in X$ has an application-dependent static weight $w(x)$.
Given a query value $s$, we say that $x$ is stabbed by $s$ iff $x.l \leq s \leq x.r$.
For ease of presentation, we first define the stabbing query:

\begin{definition}[Stabbing query]  \label{definition:stabbing}
Given a stabbing query $s$ (which is a value) and $X$, this query retrieves a subset $X_{s}$ of $X$ such that $X_{s} = \{x \,|\, x \in X, x.l \leq s \leq x.r\}$.
\end{definition}

\noindent
This paper considers a variant of stabbing queries and addresses the problem defined below.

\begin{definition}[Top-k weighted stabbing query]
Given a top-k weighted stabbing query $q = (s,k)$, where $s$ and $k$ respectively are a query value and a result size, and $X$, this query retrieves $k$ intervals with the largest weights among $X_{s}$.
(If $|X_{s}| < k$, all intervals in $X_{s}$ are returned.)
Ties are broken arbitrarily.
\end{definition}

The state-of-the-art algorithm \cite{xu2017efficiently} requires $O(\log n + m\log k)$ time, where $m = |X_{s}|$.
Theoretically, $m$ can be as large as $n$, so it requires $O(n\log k)$ time in the worst case.
In practice, if $m$ is small, this algorithm is sufficiently fast, but it is slow when $m$ is large.
We solve this issue, and the objective of this paper is to design exact algorithms that run in time less than $O(n)$ with $\tilde{O}(n)$ space and are practically fast.
Note that this paper assumes that $X$ is static, and efficient updates for dynamic interval data are not the scope of this paper.

\subsection{Interval Tree}
We introduce the interval tree structure \cite{edelsbrunner1980dynamic}, a building block of our algorithm presented in Section \ref{sec:intervalforest}.
This structure is similar to the binary tree structure, and its height is $O(\log n)$.
Each node of an interval tree has the following:
\begin{itemize}
    \setlength{\leftskip}{-5.0mm}
    \item   $v_{cen}$: the central point.
    \item   $A_{left}$: an array consisting of all intervals $x$ such that $x.l \leq v_{cen} \leq x.r$, and the intervals are sorted in ascending order of the left endpoint.
    \item   $A_{right}$: an array consisting of the same intervals as those in $A_{left}$, and they are sorted in ascending order of the right endpoint.
    \item   A left child node, and every interval $x'$ maintained by the sub-tree rooted in this left child node guarantees that $x'.r < v_{cen}$.
    \item   A right child node, and every interval $x'$ maintained by the sub-tree rooted in this right child node guarantees that $x'.l > v_{cen}$.
\end{itemize}

\noindent
\textbf{Building.}
Given $X$, a root node is first created.
From all endpoints, $v_{cen}$ is obtained, and then $A_{left}$ and $A_{right}$ are computed by using $v_{cen}$.
After that, a left (right) node is created based on $X_{1} = \{x \,|\, x \in X, x.r < v_{cen}\}$ ($X_{2} = \{x \,|\, x \in X, x.l > v_{cen}\}$).
This partition is recursively done until we can no longer partition a given subset of $X$.

\begin{figure*}[!t]
    \begin{center}
        \subfigure[Interval tree]{%
            \includegraphics[width=0.47\linewidth]{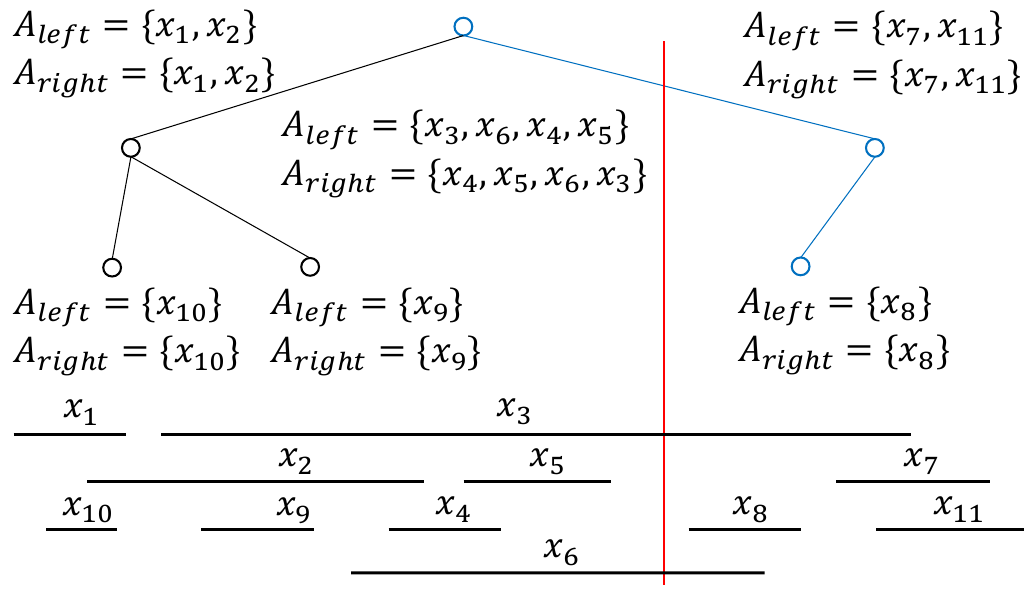} \label{fig:it}}
        \subfigure[Segment tree]{%
    	\includegraphics[width=0.47\linewidth]{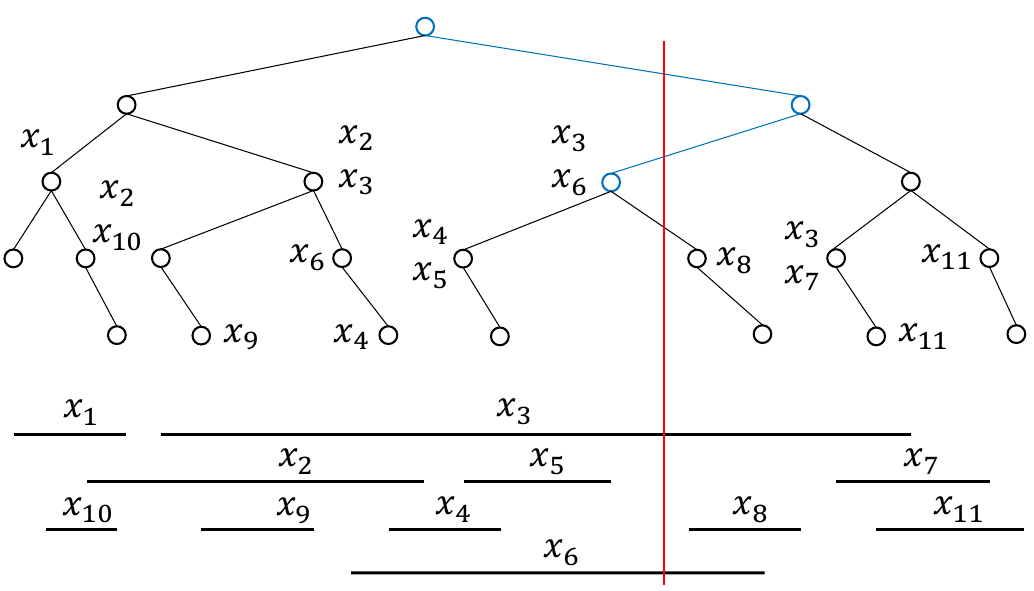}      \label{fig:st}}
        \caption{Example of the interval and segment tree structures.
        The red line represents a simple stabbing query $s$, and the traversed path is blue.
        Note that $x_{3}$ and $x_{6}$ are stabbed by the query.}
        \label{fig:structure}
    \end{center}
\end{figure*}

\vs
\noindent
\textbf{Stabbing query.}
Consider that we are given a simple stabbing query $s$ (see Definition \ref{definition:stabbing}).
We traverse the interval tree from its root.
If $s \leq v_{cen}$ ($s > v_{cen}$) of the root node, we access $A_{left}$ ($A_{right}$) and sequentially scan it until we have $s < x.l$ ($s > x.r$).
Then, we next traverse its left (right) child node (if it exists).
This is repeated until we reach a leaf node.
Fig. \ref{fig:it} illustrates an example of the interval tree structure.
The red vertical line represents a stabbing query, whereas the traversed path is blue.

\vs
\noindent
\textbf{Performance guarantee.}
The above structure and algorithm yield the following performance guarantee \cite{edelsbrunner1980dynamic}.

\begin{lemma}   \label{lemma:intervaltree}
An interval tree can be built in $O(n\log n)$ time, consumes $O(n)$ space, and processes a stabbing query in $O(\log n + m)$ time, where $m$ is the number of stabbed intervals.
\end{lemma}

\subsection{Segment Tree}   \label{sec:segmenttree}
We next introduce the segment tree structure \cite{de2000computational}, because we use it as a building block of our algorithm presented in Section \ref{sec:stpsa}.
This structure is similar to the balanced binary search tree (BST) structure, so its height is also $O(\log n)$.
However, each interval can be maintained in multiple nodes.
Each node $u$ of a segment tree has the following:
\begin{itemize}
    \setlength{\leftskip}{-5.0mm}
    \item   $key$: the endpoint of an interval.
    \item   $parent(u)$: the parent node of $u$.
    \item   $y(u)$: an interval of the minimum and maximum keys maintained by the sub-tree rooted at $u$.
    \item   $X(u)$: a set of intervals that cover $y(u)$ but are not maintained in $X(parent(u))$.
    \item   Left and right child nodes that follow the BST structure w.r.t key.
\end{itemize}

\noindent
\textbf{Building.}
We first build a balanced BST by using a set of all left and right endpoints in $X$.
Then, for each node $u$, $y(u)$ is computed in a bottom-up manner.
After that, we insert each $x \in X$ into the BST so that $x$ satisfies the above constraint.

\vs
\noindent
\textbf{Stabbing query.}
Given a stabbing query $s$, we traverse the segment tree from its root $u_{root}$.
We enumerate all intervals $\in X(u_{root})$ because all intervals in $X(u_{root})$ are guaranteed to cover $y(u_{root})$.
Then, if $s \in y$ of the left or right node of $u_{root}$, we traverse the corresponding child node.
This is repeated until we reach a leaf node or a currently accessed node has no child nodes such that $s \in y$.
Fig. \ref{fig:st} illustrates an example of the segment tree structure and the path traversed for the given stabbing query.

\vs
\noindent
\textbf{Performance guarantee.}
The segment tree structure yields the following performance guarantee \cite{de2000computational}.

\begin{lemma}   \label{lemma:segmenttree}
A segment tree can be built in $O(n\log n)$ time, consumes $O(n\log n)$ space, and processes a stabbing query in $O(\log n + m)$ time, where $m$ is the number of stabbed intervals.
\end{lemma}

\section{Algorithm based on Interval Forest} \label{sec:intervalforest}
This section proves the following theorem.

\begin{theorem} \label{theorem:if}
For our problem, there exists an exact algorithm that needs $O(n\log n)$ pre-processing time, $O(n)$ space, and $O(\sqrt{n}\log n + k)$ query time.
\end{theorem}

\noindent
\textbf{Main idea.}
The main idea of this algorithm is to combine weight-based sorting and the interval tree structure.
Assume that the intervals in $X$ are sorted in descending order of weight.
Now assume that $X$ is partitioned into two disjoint subsets $X_{1}$ and $X_{2}$, and note that $w(x) \geq w(x')$ for all $x \in X_{1}$ and $x' \in X_{2}$.
Next consider that two interval trees $\mathcal{I}_{1}$ and $\mathcal{I}_{2}$ are built, i.e., $\mathcal{I}_{1}$ ($\mathcal{I}_{2}$) is built on $X_{1}$ ($X_{2}$).
Given a top-k weighted stabbing query, we first use $\mathcal{I}_{1}$.
If $\mathcal{I}_{1}$ returns $k$ stabbed intervals, we do not need to use $\mathcal{I}_{2}$, since the weights of the intervals in $\mathcal{I}_{2}$ are less than those of the intervals in $\mathcal{I}_{1}$.
Based on this observation, we reduce the $O(n\log k)$ time of \cite{xu2017efficiently} to $O(\sqrt{n}\log n + k)$.

\subsection{Data Structure and Construction}
We sort the intervals $\in X$ as above.
Then, we partition $X$ into $p$ equal-sized disjoint subsets, i.e., $X = X_{1} \cup X_{2} \cup \cdots \cup X_{p}$ and $X_{i} \cap X_{j} = \varnothing$ ($i \neq j$).
In addition, $w(x) \geq w(x')$ for all $x \in X_{i}$, $x' \in X_{i+1}$ ($i \in [1,p-1]$).
We later show how to specify $p$, which is an important factor for achieving a solid performance guarantee.
Then, we build an interval tree for each subset of $X$, so we have $p$ interval trees.
Note that
(i) this structure is general for arbitrary top-k weighted stabbing queries, meaning that this pre-processing is done only once, and
(ii) Lemma \ref{lemma:intervaltree} directly derives the following.

\begin{corollary}   \label{corollary:if}
We can build $p$ interval trees in $O(n \log n)$ time, and they require $O(n)$ space in total.
\end{corollary}

\begin{algorithm}[!t]
    \caption{\textsf{IF} (Interval Forest algorithm)}	\label{algo:if}
    \DontPrintSemicolon
    \KwIn {$X$, $q = (s,k)$, and $p$ interval trees ($\mathcal{I}_{1}, ..., \mathcal{I}_{p}$)}
    \KwOut{$R$ (top-k result)}
    $R \gets \varnothing$   \Comment*[r]{\scriptsize initialize the top-k result $R$}
    \ForEach {$i \in [1,p]$}
    {
        $R \gets$ \textsc{Stabbing}$(\mathcal{I}_{i},q,R)$  \Comment*[r]{\scriptsize update $R$ from the stabbed intervals} \label{algo:if:stabbing}
        \textbf{If} {$|R| = k$} \textbf{then} \textbf{return} $R$
    }
    \textbf{return} $R$
\end{algorithm}

\subsection{Query Processing Algorithm}
Algorithm \ref{algo:if} describes our algorithm proposed in this section, which is denoted by \textsf{IF} (because this algorithm employs multiple interval trees, i.e., Interval Forest).
Given a top-k weighted stabbing query $q$, \textsf{IF} first uses the interval tree $\mathcal{I}_{1}$ on $X_{1}$ and runs $q$ on $\mathcal{I}_{1}$.
\textsf{IF} uses the stabbing query processing algorithm on the interval tree structure to find stabbed intervals.
Whenever \textsf{IF} accesses a stabbed interval, it updates the top-k result.
(Line \ref{algo:if:stabbing} represents these procedures.)
If the number of stabbed intervals is equal to or more than $k$, it is guaranteed that we can obtain the exact top-k result from $\mathcal{I}_{1}$, so \textsf{IF} returns the result.
Otherwise, \textsf{IF} runs $q$ on $\mathcal{I}_{2}$, and \textsf{IF} repeats this iteration until we have $k$ stabbed intervals or all interval trees are used.

\vs
\noindent
\textbf{Analysis.}
We set $p = O(\sqrt{n})$, so we have $O(\sqrt{n})$ interval trees and $|X_{i}| = O(\sqrt{n})$ for each $i \in [1,p]$.
Then, we have:

\begin{lemma} \label{lemma:if}
Algorithm \ref{algo:if} runs in $O(\sqrt{n}\log n + k)$ time.
\end{lemma}

\noindent
\textsc{Proof.}
Clearly, the worst case is to access all interval trees $\mathcal{I}_{1}, ..., \mathcal{I}_{p}$.
Let $k_{i}$ be the number of stabbed intervals obtained from $X_{i}$, and in the above case, we have $\sum_{i=1}^{p-1}k_{i} < k$.
Now consider the worst case: after running $q$ on $\mathcal{I}_{p}$, we obtain $O(\sqrt{n})$ stabbed intervals, i.e., $q$ stabs all intervals in $X_{p}$.
From Lemma \ref{lemma:intervaltree}, the time required for this case is 
\begin{eqnarray*}
    O(\log n^{1/2} + k_1) + \cdots + O(\log n^{1/2} + k_{p-1}) + O(\log n^{1/2} + \sqrt{n}\log k) \\
    = O(\sqrt{n}\log n + \sum_{i=1}^{p-1}k_i + \sqrt{n}\log k) = O(\sqrt{n}\log n + k),
\end{eqnarray*}
so this lemma holds. \wsq

\vs
\noindent
\textsc{Proof of Theorem \ref{theorem:if}.}
From Corollary \ref{corollary:if} and Lemma \ref{lemma:if}.
\wsq

\vs
\noindent
\textsc{Remark 1.}
Theorem \ref{theorem:if} proves that Algorithm \ref{algo:if} theoretically outperforms the state-of-the-art algorithm \cite{xu2017efficiently}.
In addition, this result proves that we can obtain the exact result without accessing $n$ intervals (by assuming that $k = O(1)$).
In a practical view, Algorithm \ref{algo:if} usually accesses much less than $p$ interval trees.
This means that, different from the state-of-the-art \cite{xu2017efficiently}, Algorithm \ref{algo:if} can prune unnecessary stabbed intervals.

\section{Algorithm based on a Variant of Segment Tree}   \label{sec:stpsa}
We next consider accelerating the search efficiency further (by sacrificing pre-processing time and the space complexity a bit) and prove that

\begin{theorem} \label{theorem:stpsa}
For our problem, there exists an exact algorithm that requires $O(n \log n\log\log n)$ pre-processing time, $O(n\log^{2}n)$ space, and $O(\log n + k)$ query time.
\end{theorem}

\noindent
\textbf{Main idea.}
This algorithm is designed based on the segment tree structure.
One may come up with the idea of sorting the intervals maintained in each node of a segment tree based on weight.
This idea enables access to at most $k$ intervals for each traversed node, as can be seen from Fig. \ref{fig:st}.
As the height of the segment tree is $O(\log n)$, this idea derives an $O(k\log n)$ time algorithm.
Although this algorithm can theoretically be faster than Algorithm \ref{algo:if}, its running time can be sensitive to $k$.
We therefore do not employ this approach.

Instead, we focus on the following property:
the stabbing query algorithm on the segment tree structure exploits the fact that the intervals maintained in the traversed nodes are guaranteed to be stabbed by a given query (see Section \ref{sec:segmenttree}). 
Then, by storing all intervals existing in the path from the root to each node in a sorted array, we do not need to enumerate $k$ intervals for each traversed node\footnote{This idea is not available for the interval tree structure.
This is because the interval tree structure does not guarantee that all intervals maintained in a node are stabbed by a given query.}.
This new idea and the path-based auxiliary structure are specific to our problem, since simple stabbing queries enumerate all stabbed intervals and do not consider weights.

\subsection{Variant of Segment Tree and Its Construction}
We first build a segment tree on $X$.
Then, for each node $u$ of the segment tree, we consider the path from $u_{root}$ to $u$.
We collect all ``distinct'' intervals maintained in the nodes on the path (since duplicate intervals may exist in the path), and $u$ stores this set of intervals in a weight-based sorted array.

\begin{example}
\textit{In Fig. \ref{fig:st}, assume that the blue path consists of nodes $u_{root}$, $u_{2}$, and $u_{5}$, where $u_{5}$ maintains $x_{3}$ and $x_{6}$.
Then, $u_{root}$ and $u_{2}$ do not maintain any intervals in their sorted arrays because there exist no intervals on the paths from $u_{root}$ to them.
On the other hand, assuming $w(x_{6}) > w(x_{3})$, $u_{5}$ maintains $x_{6}$ and $x_{3}$ in its sorted array in this order.}
\end{example}

After making this sorted array for each node $u$ of the segment tree, we remove $X(u)$ (a set of intervals initially maintained in $u$) because we do not use it anymore.
It can be seen that, compared with the original segment tree structure, our data structure replaces $X(u)$ with the sorted array.
Note that this structure is also general to arbitrary top-k weighted stabbing queries, so this pre-processing is done only once.
We analyze this pre-processing time and the space complexity of this structure.

\begin{lemma}   \label{lemma:segpta-preprocess}
We need $O(n\log n\log\log n)$ time to build the above variant of a segment tree.
\end{lemma}

\noindent
\textsc{Proof.}
From Lemma \ref{lemma:segmenttree}, we can build a segment tree in $O(n\log n)$ time.
In the segment tree, there exist $O(n)$ nodes, so we need to consider $O(n)$ paths.
Moreover, $O(n\log n)$ intervals exist in the segment tree.
Given these facts, we see that the amortized number of intervals in each path is $O(\log n)$.
The cost of sorting these intervals is $O(\log n\log\log n)$.
Therefore, the total cost of making path arrays for $O(n)$ nodes is $O(n\log n\log\log n)$.
\wsq

\begin{lemma}   \label{lemma:segpta-space}
The above variant of a segment tree needs $O(n\log^{2}n)$ space.
\end{lemma}

\noindent
\textsc{Proof.}
Recall that the original segment tree has $O(n\log n)$ intervals.
These intervals can be replicated in additional $O(\log n)$ nodes, as the length of each path is $O(\log n)$.
Now this lemma is clear.
\wsq

\vs
\noindent
\textsc{Remark 2.}
The space requirement of our new segment tree is near linear to $n$ theoretically.
However, it practically scales linearly to $n$ because each interval is rarely replicated in $O(\log n)$ nodes.
Our experimental results also demonstrate this fact, see Section \ref{sec:experiment:memory}.

\subsection{Query Processing Algorithm}
Now we are ready to present our second algorithm for the top-k weighted stabbing queries.
Thanks to our non-trivial extension of the segment tree structure, we can design a simple and fast algorithm.
This algorithm is denoted by \textsf{ST-PSA} (Segment Tree with Path-based Sorted Arrays).

Algorithm \ref{algo:stpsa} shows each step of \textsf{ST-PSA}.
Let $\mathcal{S}$ be our variant of a segment tree on $X$.
Given a top-k weighted stabbing query $q = (s,k)$, \textsf{ST-PSA} first runs a simple stabbing query $q.s$ on $\mathcal{S}$ and obtains the node traversed last during the stabbing.
Let this node be $u$, and \textsf{ST-PSA} uses the sorted array of $u$.
Specifically, \textsf{ST-PSA} returns the first $k$ intervals in the array as the top-k result.

\begin{algorithm}[!t]
    \caption{\textsf{ST-PSA} (Segment Tree with Path-based Sorted Array algorithm)}	\label{algo:stpsa}
    \DontPrintSemicolon
    \KwIn {$X$, $q = (s,k)$, and $\mathcal{S}$ (our variant of a segment tree)}
    \KwOut{$R$ (top-k result)}
    $R \gets \varnothing$   \Comment*[r]{\scriptsize initialize the top-k result $R$}
    $u \gets$ \textsc{Stabbing}$(\mathcal{S},q.s)$  \Comment*[r]{\scriptsize obtain the last traversed node of $\mathcal{S}$}   \label{algo:stpsa:stabbing}
    $R \gets$ the first $k$ intervals in the sorted array of $u$\;  \label{algo:stpsa:path}
    \textbf{return} $R$
\end{algorithm}

\vs
\noindent
\textbf{Correctness.}
Recall the stabbing query algorithm on the segment tree structure: all intervals maintained in the traversed nodes are stabbed by a given query.
In addition, the sorted array of $u$ stores all intervals (initially) maintained in the path from $u_{root}$ to $u$.
From these facts, the correctness of \textsf{ST-PSA} is clear.

\vs
\noindent
\textbf{Time complexity.}
We present the main result of this section below.

\begin{lemma} \label{lemma:stpsa:time}
Algorithm \ref{algo:stpsa} runs in $O(\log n + k)$ time.
\end{lemma}

\noindent
\textsc{Proof.}
From Lemma \ref{lemma:segmenttree}, line \ref{algo:stpsa:stabbing} needs $O(\log n)$ time.
Line \ref{algo:stpsa:path} trivially accesses at most $k$ intervals.
Therefore, this lemma holds.
\wsq

\vs
\noindent
\textsc{Proof of Theorem \ref{theorem:stpsa}.}
From Lemmas \ref{lemma:segpta-preprocess}--\ref{lemma:stpsa:time}.
\wsq


\section{Experiment}    \label{sec:experiment}
This section reports our experimental results.
All experiments were conducted on a Ubuntu 22.04 LTS machine with 2.2GHz Intel Core i9-13950HX processor and 128GB RAM.

\vs
\noindent
\textbf{Dataset.}
We used two real datasets, BTC\footnote{https://www.kaggle.com/datasets/swaptr/bitcoin-historical-data} and Renfe\footnote{https://www.kaggle.com/datasets/thegurusteam/spanish-high-speed-rail-system-ticket-pricing}.
BTC is a set of 2,538,921 historical price intervals of Bitcoin.
Low and high prices were used as the left- and right endpoints, respectively.
Renfe is a set of 38,753,060 Spanish rail trips.
We used departure time and arrival time as the left and right endpoints, respectively.
The weight of each interval in the two datasets followed a Gaussian distribution, where mean and variance were 5000 and 1500, respectively.

\vs
\noindent
\textbf{Queries.}
We generated 1,000 top-k weighted stabbing queries.
The query value of each top-k weighted stabbing query was drawn uniformly at random from the domain of a given dataset.
The default $k$ was 25.

\vs
\noindent
\textbf{Evaluated algorithms.}
We evaluated the following algorithms.
\begin{itemize}
    \setlength{\leftskip}{-5.0mm}
    \item   \textsf{SS} (Sequential Scan):
            This algorithm sorts $X$ in descending order of weight in the pre-processing phase.
            Given a top-k weighted stabbing query, it scans $X$ until $k$ stabbed intervals are found.
    \item   \textsf{IT} (Interval Tree):
            This algorithm uses an interval tree to find all stabbed intervals and, among them, it finds $k$ intervals with the largest weight.
            This algorithm is equivalent to the state-of-the-art algorithm \cite{xu2017efficiently}.
    \item   \textsf{IF}: Our algorithm presented in Section \ref{sec:intervalforest} (Algorithm \ref{algo:if}).
    \item   \textsf{ST-PSA}: Our algorithm presented in Section \ref{sec:stpsa} (Algorithm \ref{algo:stpsa}).
\end{itemize}
The above algorithms were single-threaded, implemented in C++, and compiled by g++ 11.3.0 with -O3 flag.

\subsection{Pre-processing Time}    \label{sec:experiment:preprocessing}
We first investigated the pre-processing times of \textsf{IT}, \textsf{IF}, and \textsf{ST-PSA}.
(As \textsf{SS} requires only a single sorting and does not build any data structures, we do not discuss its pre-processing time.)
The result is shown in Table \ref{tab:preprocessing}.
\textsf{IT} and \textsf{IF} can be built faster than \textsf{ST-PSA} on BTC, but, on Renfe, they show similar pre-processing times.
This result implies that the pre-processing time of each algorithm depends on the distribution of a given dataset.
The result of similar pre-processing times of \textsf{IT} and \textsf{IF} is reasonable, as analyzed theoretically in Section \ref{sec:intervalforest}.
The result in Table \ref{tab:preprocessing} suggests that each data structure can be built in a reasonable time.

\begin{figure*}[!t]
    \begin{center}
        \subfigure[BTC]{%
            \includegraphics[width=0.30\linewidth]{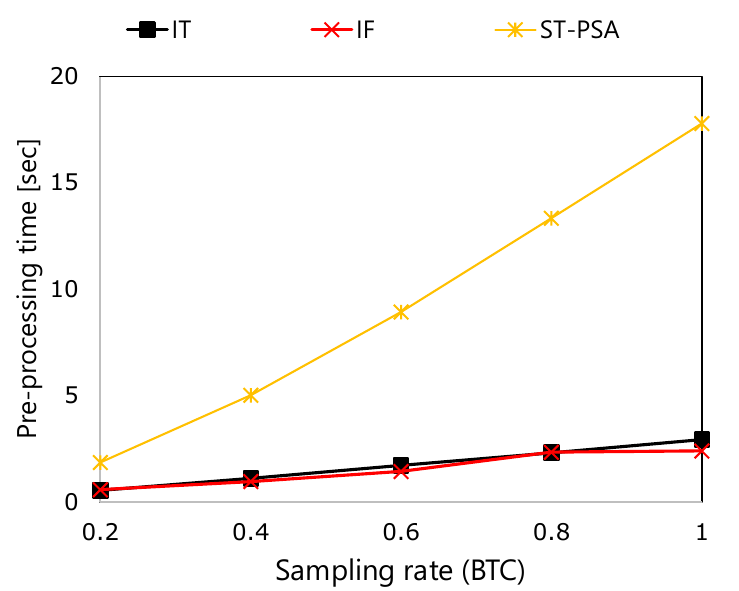}   \label{fig:preprocessing-btc}}
        \subfigure[Renfe]{%
    	\includegraphics[width=0.30\linewidth]{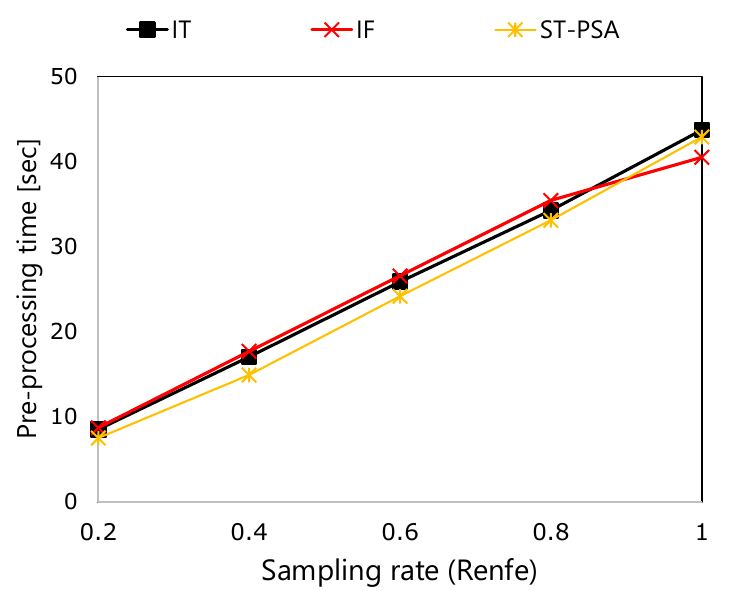} \label{fig:preprocessing-renfe}}
        \caption{Pre-processing time [sec] vs. dataset size}
        \label{fig:preprocessing}
    \end{center}
\end{figure*}

To study the impact of data size on the pre-processing times of \textsf{IT}, \textsf{IF}, and \textsf{ST-PSA}, we randomly sampled intervals in $X$ with probability of a certain sampling rate and varied this rate.
Fig. \ref{fig:preprocessing} shows the result.
Although the construction times of these structures need near linear time w.r.t. $n$ theoretically, they practically scale linearly to $n$ (except in the case of \textsf{ST-PSA} on BTC).

\begin{table}[!t]
    \centering
    \caption{Pre-processing time [sec]}
    \label{tab:preprocessing}
    \begin{tabular}{cccc}    \toprule
        Dataset & \textsf{IT}   & \textsf{IF}   & \textsf{ST-PSA}   \\ \midrule
        BTC     & 2.93          & 2.40          & 17.80             \\ 
        Renfe   & 43.71         & 40.52         & 42.91             \\ \bottomrule
    \end{tabular}
\end{table}

\begin{figure*}[!t]
    \begin{center}
        \subfigure[BTC]{%
            \includegraphics[width=0.30\linewidth]{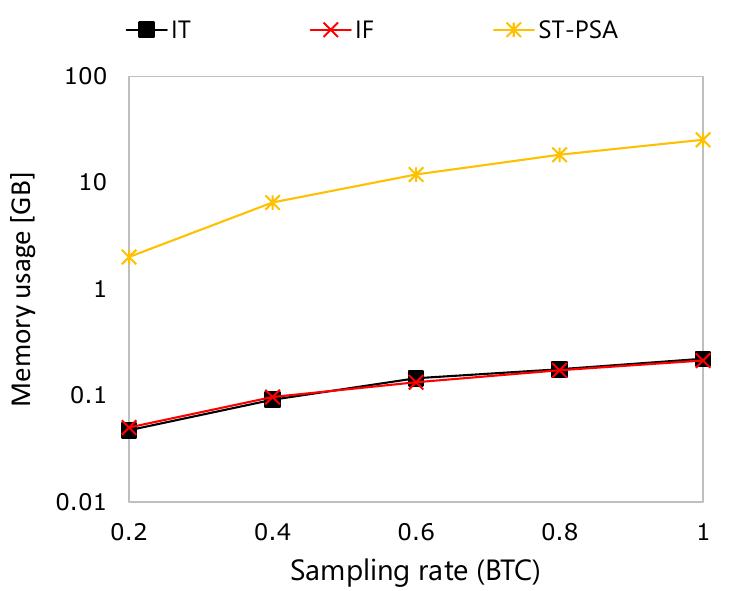}   \label{fig:memory-btc}}
        \subfigure[Renfe]{%
    	\includegraphics[width=0.30\linewidth]{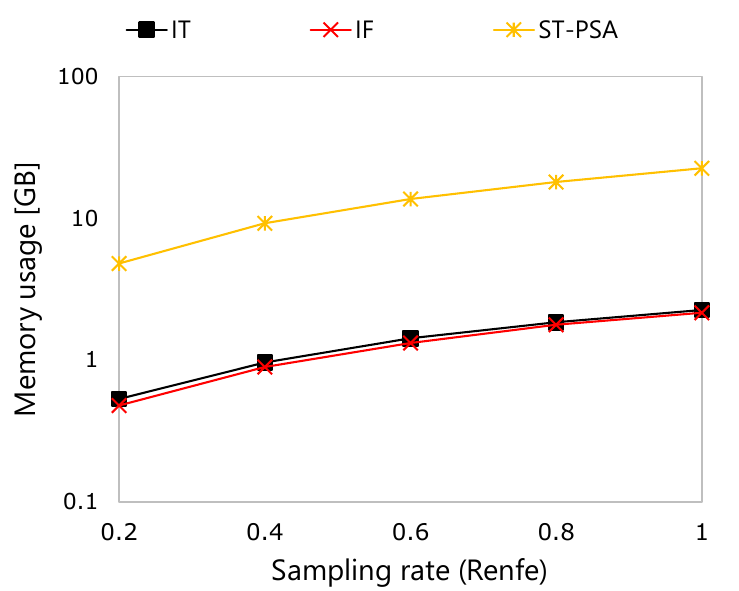} \label{fig:memory-renfe}}
        \caption{Memory usage [MB] vs. dataset size}
        \label{fig:memory}
    \end{center}
\end{figure*}

\subsection{Memory Usage}   \label{sec:experiment:memory}
Next, we focus on the memory usages of \textsf{IT}, \textsf{IF}, and \textsf{ST-PSA}.
(Recall that \textsf{SS} does not require additional spaces.)
Recall that \textsf{IT} and \textsf{IF} need $O(n)$ space, whereas \textsf{ST-PSA} requires $O(n\log^{2}n)$ space.
Table \ref{tab:memory} shows the result, and, as with the theoretical result, \textsf{ST-PSA} requires more memory than the others.
Although the memory usage of \textsf{ST-PSA} is several dozen gigabytes (on million-scale datasets), this is affordable for modern servers, as they often have terabyte-scale RAM \cite{zhang2015memory}.
The memory usages of \textsf{IT} and \textsf{IF} are similar, which is also reasonable, because the number of nodes can be almost the same.

Similar to the pre-processing time experiments, we studied the impact of data size on memory usage.
Fig. \ref{fig:memory} describes the result.
In practice, the space of \textsf{ST-PSA} scales linearly to $n$ rather than $O(n\log^{2}n)$.

\begin{table}[!t]
    \centering
    \caption{Memory usage [GB]}
    \label{tab:memory}
    \begin{tabular}{cccc}  \toprule
        Dataset & \textsf{IT}   & \textsf{IF}   & \textsf{ST-PSA}   \\ \midrule
        BTC     & 0.22          & 0.21          & 24.45             \\ 
        Renfe   & 2.26          & 2.16          & 22.56             \\ \bottomrule
    \end{tabular}
\end{table}

\subsection{Query Processing Time}
We turn our attention to query processing time.
Recall that \textsf{SS}, \textsf{IT}, \textsf{IF}, and \textsf{ST-PSA} require respectively $O(n\log k)$, $O(\log n + m\log k)$ ($m$ is the number of stabbed intervals), $O(\sqrt{n}\log n + k)$, and $O(\log n + k)$ times.
It is important to note that BTC has a small $m$, while Renfe has a large $m$.
This setting is useful to compare the performances of our algorithms with that of \textsf{IT}, as it can be fast/slow on BTC/Renfe.

\vs
\noindent
\textbf{Ablation study of \textsf{ST-PSA}.}
Since \textsf{ST-PSA} uses the segment tree structure as its building block, we first compare its performance with those of the original segment tree and sorted segment tree (the intervals in each node are sorted based on weight).
This sorted segment tree supports $O(k\log n)$ time top-k weighted stabbing queries.
Table \ref{tab:ablation} exhibits the ablation study result.

We see that \textsf{ST-PSA} shows the best performance, whereas segment tree shows the worst one.
Particularly, \textsf{ST-PSA} is more than 10 times faster than segment tree.
Also, \textsf{ST-PSA} is at least two times faster than sorted segment tree.
The time complexities of these algorithms have already shown the theoretical superiority of \textsf{ST-PSA}, and this empirical result also demonstrates that our path-based approach is more appropriate than the simple modification of the segment tree structure.

\begin{table}[!t]
    \centering
    \caption{Query processing time [microsec]}
    \label{tab:ablation}
    \begin{tabular}{cccc}    \toprule
        Dataset & Segment tree  & Sorted segment tree   & \textsf{ST-PSA}   \\ \midrule
        BTC     & 8.94          & 6.65                  & 1.76              \\ 
        Renfe   & 14.34         & 3.24                  & 1.23              \\ \bottomrule
    \end{tabular}
\end{table}

\begin{figure*}[!t]
    \begin{center}
        \subfigure[BTC]{%
            \includegraphics[width=0.30\linewidth]{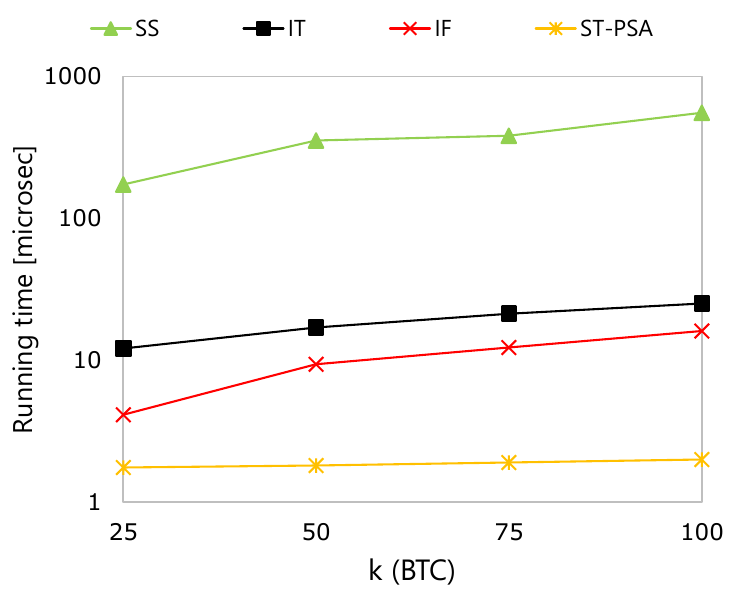}   \label{fig:k-btc}}
        \subfigure[Renfe]{%
    	\includegraphics[width=0.30\linewidth]{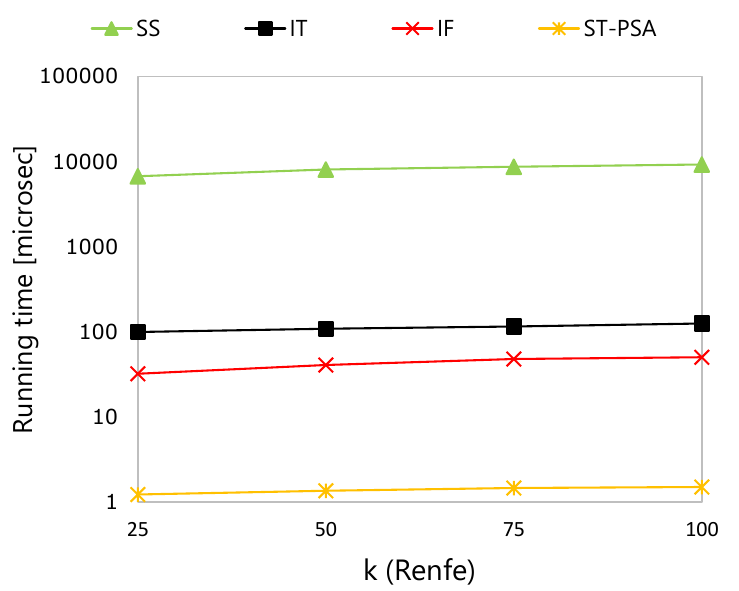} \label{fig:k-renfe}}
        \caption{Running time vs. $k$}
        \label{fig:k}
    \end{center}
\end{figure*}
\begin{figure*}[!t]
    \begin{center}
        \subfigure[BTC]{%
            \includegraphics[width=0.30\linewidth]{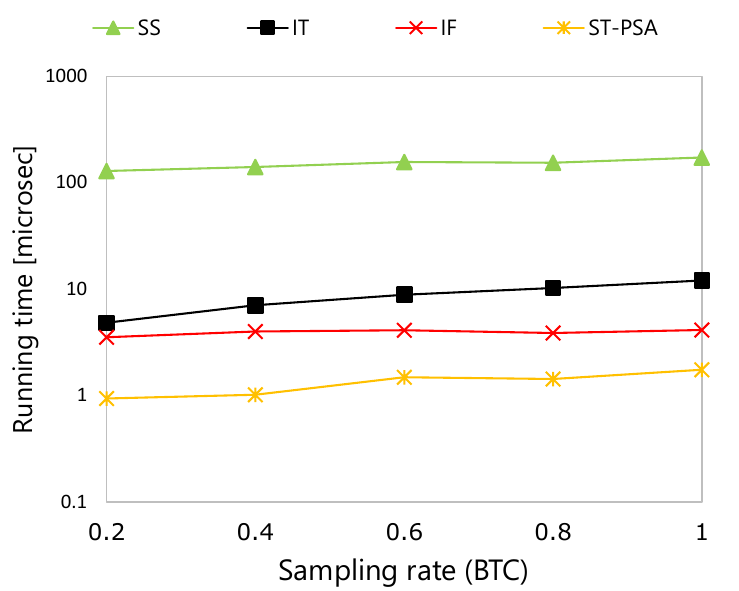}   \label{fig:cardinality-btc}}
        \subfigure[Renfe]{%
    	\includegraphics[width=0.30\linewidth]{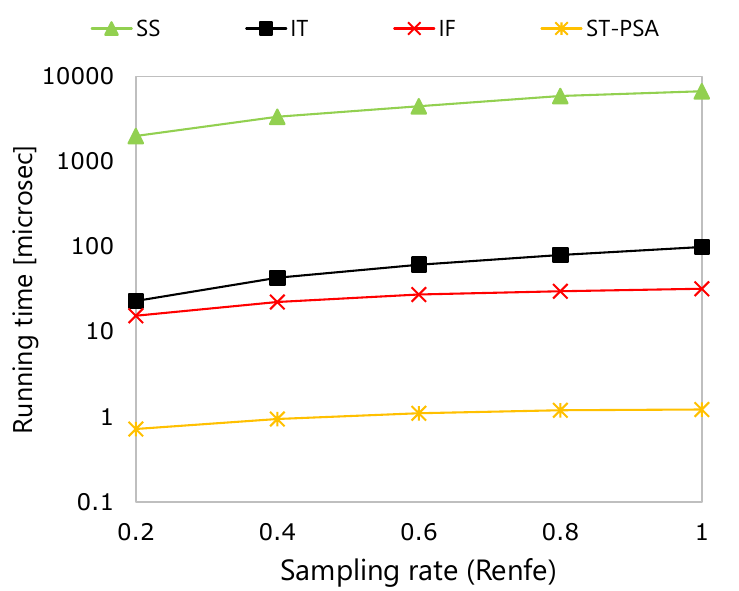} \label{fig:cardinality-renfe}}
        \caption{Running time vs. data size}
        \label{fig:cardinality}
    \end{center}
\end{figure*}

\vs
\noindent
\textbf{Impact of $k$.}
We next compare our algorithms with \textsf{SS} and \textsf{IT} by varying $k$.
Fig. \ref{fig:k} shows the experimental result.
Our algorithms outperform \textsf{SS} and \textsf{IT} on BTC and Renfe.
For example, \textsf{ST-PSA} is about 5500 (80) times faster than \textsf{SS} (\textsf{IT}) on Renfe when $k = 25$.

\vs
\noindent
$\bullet\,$ \textsf{IF} vs. \textsf{IT}.
Recall that BTC has a small $m$, and even in this case, \textsf{IF} is faster than \textsf{IT}.
That is, our combination of weight-based sorting and the interval tree structure functions better than simply employing an interval tree.
This observation suggests that \textsf{IF} prunes many unnecessary stabbed intervals.

\vs
\noindent
$\bullet\,$ \textsf{IF} vs. \textsf{ST-PSA}.
Next, we see that \textsf{ST-PSA} is consistently faster than \textsf{IF}.
This result is consistent with Theorems \ref{theorem:if} and \ref{theorem:stpsa}.
In addition, \textsf{ST-PSA} is more robust than \textsf{IF} against $k$.
As $k$ increases, \textsf{IF} tends to access more interval trees, i.e., the number of stabbing operations increases.
On the other hand, \textsf{ST-PSA} accesses at most $\log n$ nodes and $k$ intervals, so it does not suffer from accessing ``more nodes'' even when $k$ increases. 

\vs
\noindent
\textbf{Impact of dataset size.}
As with the experiments in Sections \ref{sec:experiment:preprocessing} and \ref{sec:experiment:memory}, we studied the scalability to $n$ w.r.t. query processing time.
Fig. \ref{fig:cardinality} shows the result.
We have two observations.
The first one is that the query processing time of \textsf{IT} scales linearly to $n$, which demonstrates the claim of $m = O(n)$.
The other is that our algorithms scale better than \textsf{IT}, consistent with the theoretical results shown in Table \ref{tab:time}.
The running times of our algorithm increase only slightly, even when $n$ increases.
This empirical result confirms the importance of designing less than $O(n)$ time algorithms.
Our algorithms achieve this main objective.

\section{Related Work}  \label{sec:related-work}
\textbf{Stabbing queries.}
Stabbing queries return all stabbed intervals and have been studied for years because they are one of primitive operators for intervals.
The two most representative data structures for efficient stabbing query processing are the interval tree and the segment tree.
They yield $O(\log n + m)$ time algorithms, where $m$ is the number of stabbed intervals.
Since $m = O(n)$, simply applying these algorithms cannot efficiently solve our problem.
The state-of-the-art algorithm \cite{xu2017efficiently} suffers from this issue, since it simply employs the interval tree.
Although one of our algorithms also employs the interval tree structure, it exploits this structure in a more efficient way, leading to a better time complexity.
Another algorithm extends the segment tree structure in a non-trivial way, and it can prune all unnecessary intervals, as proved in Theorem \ref{theorem:stpsa}.

Some works \cite{agarwal2005optimal,kaplan2003dynamic} considered stabbing max queries on weighted intervals, and a stabbing max query finds the interval with the largest weight among a set of stabbed intervals.
This query is a special case of our problem, as it is a top-1 weighted stabbing query.
Unfortunately, these works do not consider the top-k version, and how to extend their algorithms for our problem is not trivial.

\vs
\noindent
\textbf{Range queries.}
A range query on interval data specifies an interval as a query and returns all intervals overlapping the query interval.
The problem of processing range queries on intervals has also been studied \cite{amagata2024independent,amagata2024independent_,behrend2019period,christodoulou2022hint,christodoulou2023hint,kaufmann2013timeline}.
The timeline index \cite{kaufmann2013timeline} is implemented in SAP-HANA \cite{farber2012sap}.
This index employs endpoint-based management like the interval tree structure but does not use a hierarchical structure.
The period index \cite{behrend2019period} is a hierarchical one-dimensional grid, where each hierarchy has a different grid granularity.
HINT \cite{christodoulou2022hint,christodoulou2023hint} is a state-of-the-art hierarchical index for range queries on interval data.
This structure stores intervals to adapt to the distribution of a given dataset and exploits hardware optimization.

The main drawback of these structures is that they have no interesting theoretical bound for range queries.
Thus, the query times of these techniques are $O(n)$ in the worst case.
In addition, they do not consider weighted intervals, suggesting that they are not appropriate for our problem.


\section{Conclusion}    \label{sec:conclusion}
This paper addressed the problem of processing top-k weighted stabbing queries.
A state-of-the-art algorithm for this problem incurs the same time complexity as that of a sequential scan.
Motivated by this inefficiency issue, this paper proposed two algorithms.
One runs in $O(\sqrt{n}\log n + k)$ time, and the other runs in $O(\log n + k)$ time, showing that the query times of our algorithms theoretically beat that of the state-of-the-art algorithm.
We conducted extensive experiments, and the results demonstrate that our algorithms outperform the state-of-the-art algorithm not only theoretically but also empirically.

This paper focused on static datasets.
An interesting future direction is to consider dynamic intervals and continuous top-k weighted stabbing queries.

\begin{acks}
This work was partially supported by AIP Acceleration Research JPMJCR23U2, JST, and JSPS KAKENHI Grant Number 24K14961.
\end{acks}

\bibliographystyle{ACM-Reference-Format}
\bibliography{bibtex}

\end{document}